\documentclass{PoS}

\title{Using vo tools to iNvestIgate Quasar Spectra (UNIQS)}

\ShortTitle{Using vo tools to iNvestIgate Quasar Spectra (UNIQS)}

\author{\speaker{Swayamtrupta Panda}$^{1,4}$, {Kasia Ma\l{}ek}$^{3,5}$, {Marzena {\'S}niegowska}$^{1,2}$, {Bo{\.z}ena Czerny}$^{1}$\\
        $^{1}$Center for Theoretical Physics, al. Lotnik{\'o}w 32/46, 02-668 Warsaw\\
        $^{2}$Warsaw University Observatory, al. Ujazdowskie 4, 00-478 Warsaw\\
        $^{3}$National Center for Nuclear Research, ul. Ho{\.z}a 69, 00-681 Warsaw\\
        $^{4}$Nicolaus Copernicus Astronomical Center, ul. Bartycka 18, 00-716 Warsaw\\
        $^{5}$Laboratoire d'Astrophysique de Marseille, OAMP, Universit{\'e} Aix-Marseille, CNRS, 38 rue Fr{\'e}d{\'e}ric Joliot-Curie, 13388, Marseille Cedex 13, France\\
        E-mail: \email{panda@cft.edu.pl}}

\abstract{The work initially started as a test to retrace the Quasar Main Sequence diagram $\cite{shen2}$ using \href{http://www.star.bris.ac.uk/~mbt/topcat/}{TOPCAT}, where they (and references therein) claimed that the parameter R$\mathrm{_{FeII}}$, which defines the Eigenvector 1 (EV1) is driven by the Eddington ratio alone. We subsequently construct a refined (error and redshift limited) sample from the original QSO catalog $\cite{shen1}$. Based on our hypothesis - the main driver of the Quasar Main Sequence is the  maximum of the accretion disk temperature (T$\mathrm{_{BBB}}$) defined by the Big Blue Bump on the Spectral Energy Distribution. We select five extreme sources that have R$\mathrm{_{FeII}} \geq $ 4.0 and use the SED modelling code \href{https://cigale.lam.fr/}{CIGALE} to fit the multi-band photometric data obtained using \href{http://vizier.u-strasbg.fr/vizier/sed/}{Vizier photometric viewer} and available in literature for these sources. Incorporating the prescription for broad emission-line species dependent virial factors, we derive the black hole masses for the entire QSO catalog and compare their distribution with the black hole masses in the original catalog. Here, we show the results of our analysis for one of the 5 extreme sources, SDSSJ082358.30+213545.2. We also show the detailed modeling, including the Fe II pseudo-continuum to estimate and compare the value of R$\mathrm{_{FeII}}$ for this object.}

\FullConference{Revisiting narrow-line Seyfert 1 galaxies and their place in the Universe - NLS1 Padova\\
		9-13 April 2018 \\
		Padova Botanical Garden, Italy}

\usepackage{subfig}
\usepackage{ragged2e}
\usepackage{soul}
\begin{document}

\section*{A Quasar Main Sequence diagram}
\vspace{-0.25cm}
Quasars are rapidly accreting supermassive black holes at the centres of massive galaxies. Their dependence on a broad range of parameters, many of which are highly correlated, across all wavelengths reflects the diversity in the physical conditions of the regions in the vicinity of the central core. This problem is analogous to the problem of identifying what governs the stellar main sequence on the Hertzsprung-Russel diagram $\cite{sul00}$ when the classification was based purely on spectral properties of the stellar atmospheres or plots were done in color-color diagrams.
\begin{figure}
\includegraphics[width=\textwidth]{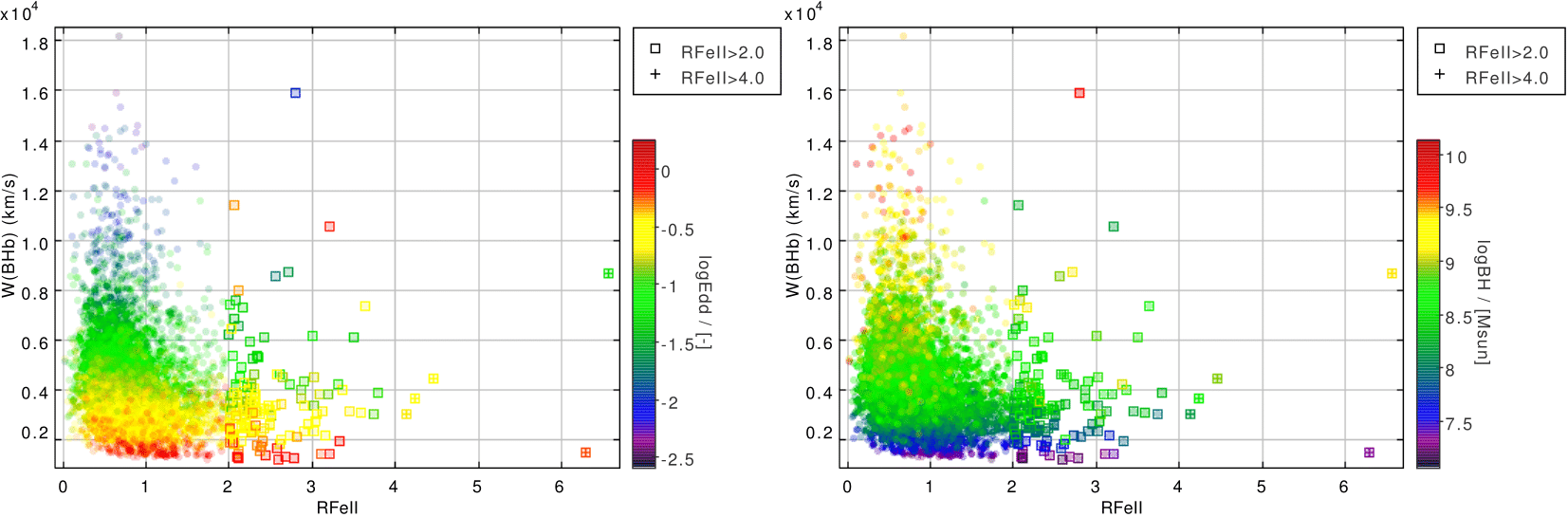}
\caption{\label{fig1} $\mathrm{FWHM(H\beta)}$ - $\mathrm{R_{FeII}}$ distributions for error-corrected and z-limited sample of 4989 quasars with auxiliary axis (i) Eddington ratio; and (ii) Black Hole Mass. The filled circles are all the quasars in the sample, the open squares is the subset with R$\mathrm{_{FeII}}$ $\geq$ 2 (157 sources) and the pluses is the subset with R$\mathrm{_{FeII}}$ $\geq$ 4 (5 sources)}
\end{figure}

\section*{The test case: SDSSJ082358.30+213545.2}
\vspace{-0.25cm}
A $\mathrm{FWHM(H\beta)} - \mathrm{R_{FeII}}$ diagram for $\sim$ 20,000 SDSS quasars was presented in $\cite{shen2}$ using the QSO catalog $\cite{shen1}$, where the relative Fe$_{\mathrm{{II}}}$ strength (R$_{\mathrm{Fe_{II}}}$= EW$_{\mathrm{Fe_{II}}}$/EW$_{\mathrm{H\beta}}$) is considered for the FeII emission within 4434-4684 \AA\,$\cite{bor}$. To remove line widths with large errors, we limit the catalog by selecting only objects with \,$<$20\% error in the FWHMs and the EWs using TOPCAT $\cite{tay}$. This gives us our base sample with 4989 quasars. Selecting sub-samples with R$_{\mathrm{Fe_{II}}}$ $\geq$ 4.0, gives us 5 sources with the maximum R$_{\mathrm{Fe_{II}}}$ = 6.56 which is almost a factor 3  more than the maximum shown in Fig.1 of $\cite{shen2}$.
\begin{figure}%
\centering
    \subfloat{{\includegraphics[width=6cm, height = 5cm]{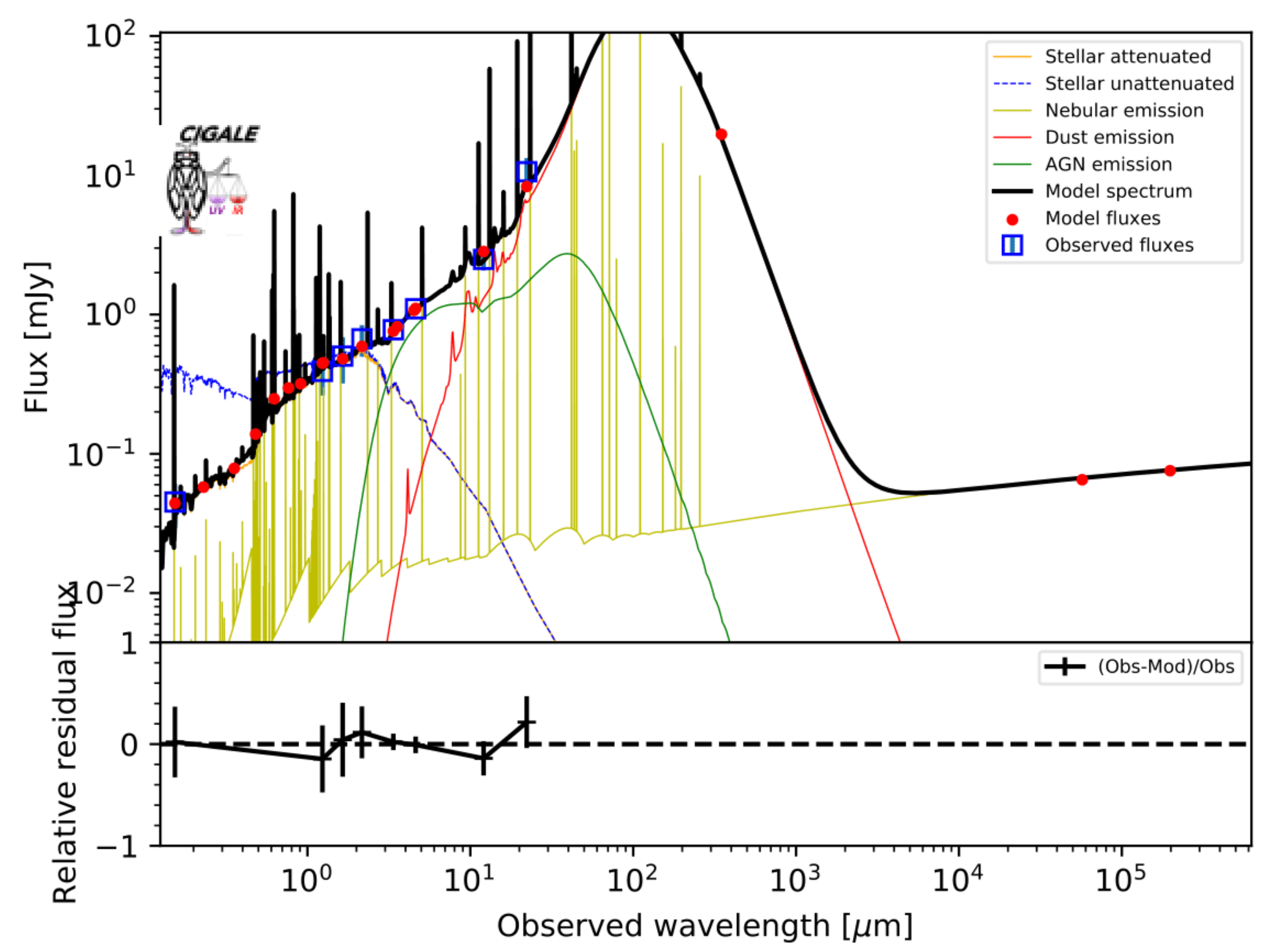} }}%
    \qquad
    \subfloat{{\includegraphics[width=6cm, height = 5cm]{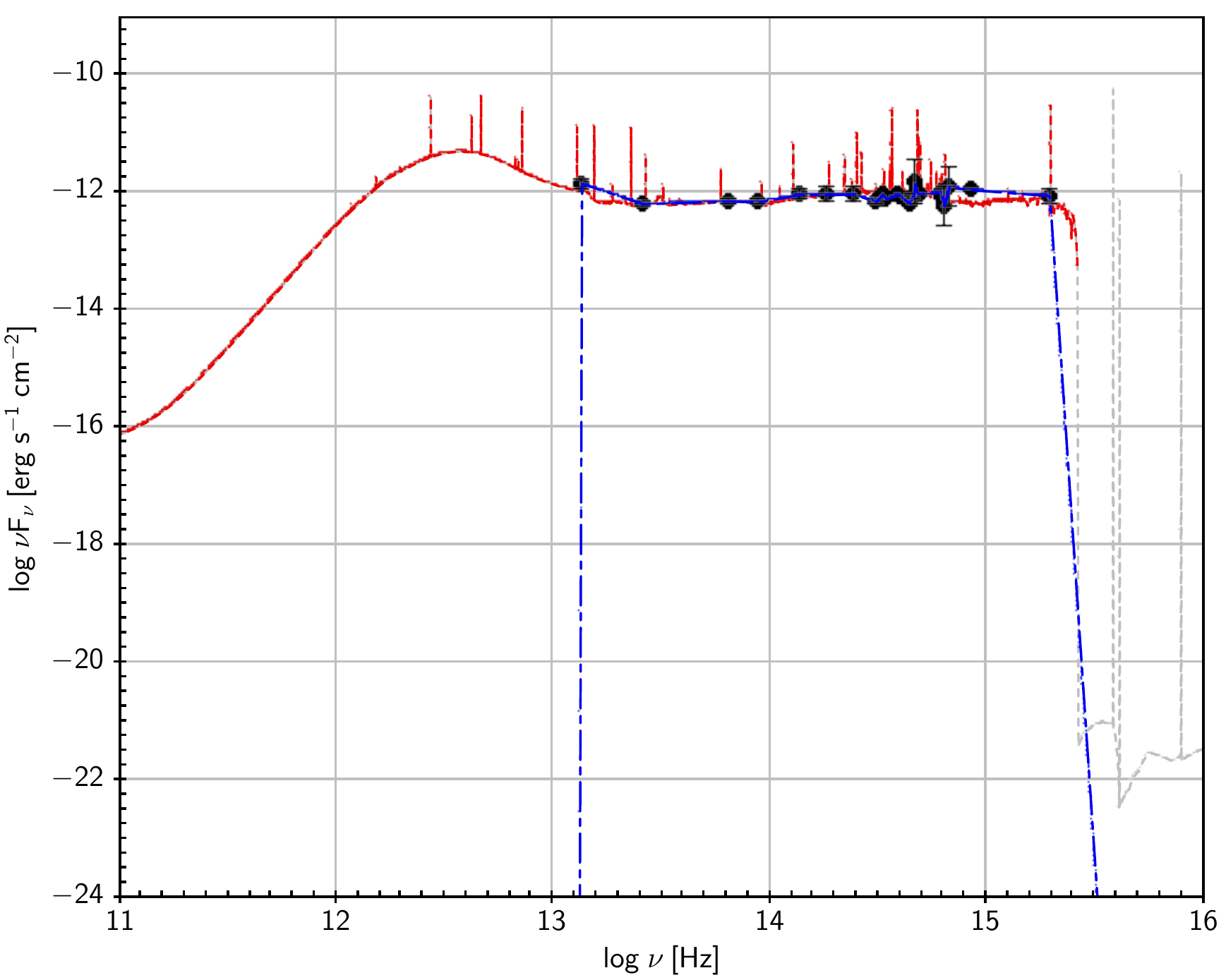} }}%
    \caption{(a) CIGALE $\cite{nol}$ best model fitting for SDSSJ082358.30+213545.20 at z = 0.248. Reduced $\chi^2$ = 0.86. The lower panel shows the relative residual flux; (b) The CIGALE model SED: The grey curve shows the total SED from the best fit. The red curve is the total (line+continuum) emission beyond the Lyman break ($\sim$912 \AA) removed. The black circles are the photometric data points with errorbars. The fit in the blue is the interpolation from CLOUDY $\cite{fer}$.}%
    \label{fig2}%
\end{figure}

\section*{Maximum of the Big Blue Bump}
\vspace{-0.25cm}
In $\cite{pan}$, we postulated that the true driver behind the EV1 is the maximum of the temperature in a multicolor accretion disk which is also the basic parameter determining the broad band shape of the quasar continuum emission:\\
\begin{equation}
    T_{BBB} = 1.732\times 10^{19} \left(\frac{\dot{M}}{M^{2}}\right)^{0.25}
    \label{eq1}
\end{equation}
\par 
We expect that this maximum temperature depends on the ratio of the Eddington ratio to the black hole mass (or, equivalently, on the ratio of the accretion rate to square of the black hole mass). The dependence of the quasar main sequence on Eddington ratio and the black hole mass is clearly seen from the Fig.\ref{fig1}. Using multi-wavelength photometric data, we produce the best fit for this source using CIGALE (Fig.\ref{fig2}a). To construct the SED, we used \href{http://vizier.u-strasbg.fr/vizier/sed/}{Vizier photometric viewer} and available literature to procure the data points. The SED is shown in Fig.\ref{fig2}b along with a CLOUDY interpolation on the overplotted data with respective errorbars.\\
\par 
\vspace{-0.15cm}
\fbox{\begin{minipage}{35em}
\begin{tiny}
\href{https://cigale.lam.fr/documentation/}{Description of the CIGALE parameters used}:
\vspace{-0.15cm}
\begin{itemize}
\justifying
\item \textit{sfhdelayedplusExpburst} - This module implements a star formation history (SFH) described as a delayed rise of the SFR up to a maximum, first followed by an exponential decrease; then followed by an exponential starburst activity.
\vspace{-0.25cm}
\item  \textit{bc03} - This module implements the $\cite{bru}$ Single Stellar Populations.
\vspace{-0.25cm}
\item \textit{nebular} - Module computing the nebular emission from the ultraviolet to the near-infrared. It includes both the nebular lines and the nubular continuum. It takes into account the escape fraction and the absorption by dust.
\vspace{-0.25cm}
\item \textit{dustatt$_{calzleit}$} - This module implements the $\cite{cal}$ and $\cite{lei}$ attenuation formulae, adding an UV-bump and a power law.
\vspace{-0.25cm}
\item \textit{dl2014} - This module implements the $\cite{dal}$ infra-red models.
\vspace{-0.25cm}
\item \textit{fritz2006} - $\cite{fri}$ AGN dust torus emission.
\end{itemize}
\end{tiny}
\end{minipage}}
\section*{Line dependent $\mathrm{M_{BH}}$ distribution}
\begin{figure}
\centering
\includegraphics[scale=0.5]{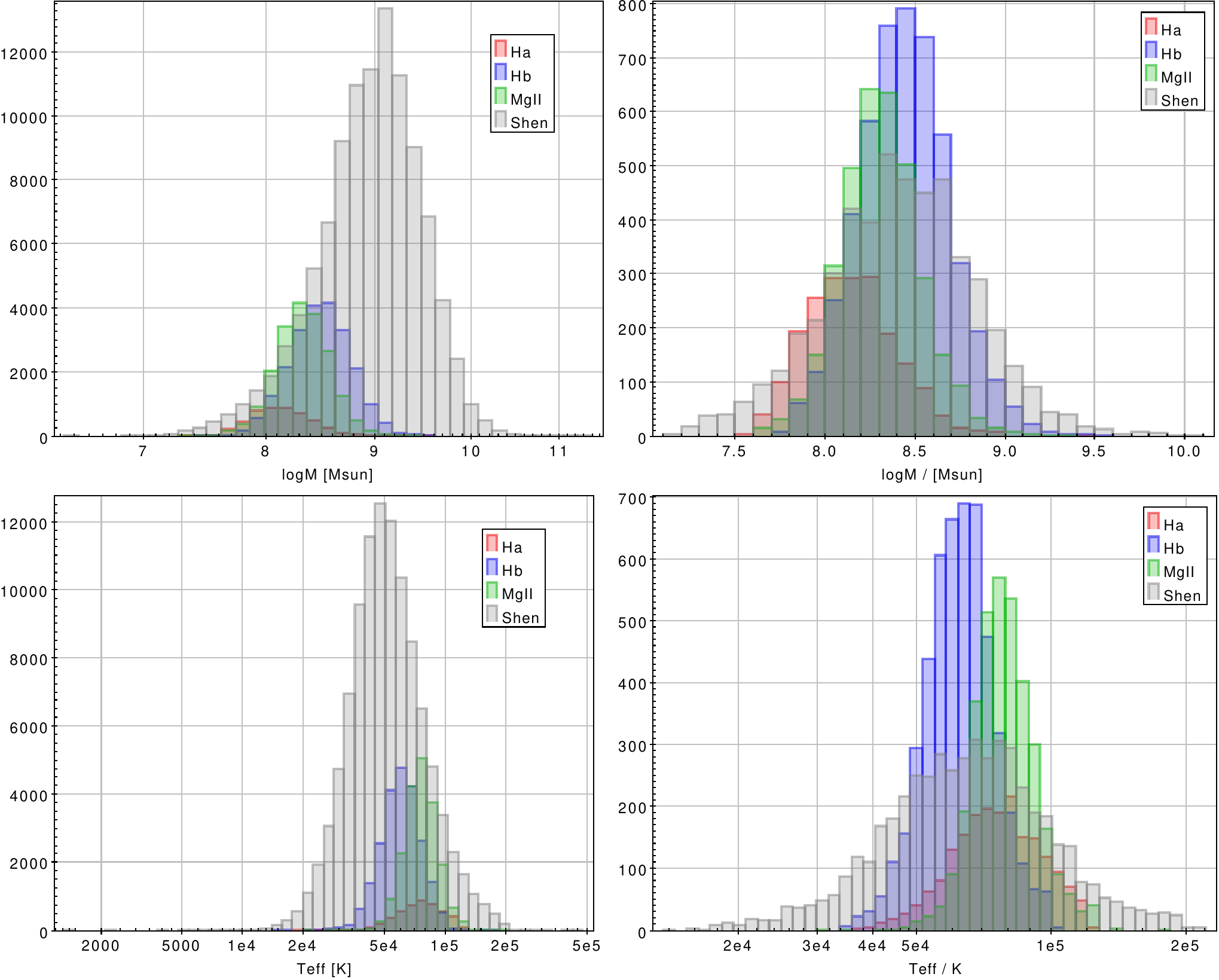}
\caption{\label{fig3}(top panel left): broad emission line-dependent $\mathrm{M_{BH}}$ distribution compared with the fiducial $\mathrm{M_{BH}}$ estimates from $\cite{shen1}$ QSO catalog for all available line widths in the catalog. (top panel right): similiar distribution for the our error-corrected sample of 4989 quasars. (bottom panel left, right): The derived T$\mathrm{_{BBB}}$ distribution for the entire catalog and for our sample.}
\end{figure}

\vspace{-0.25cm}
Taking the prescription from $\cite{mej}$, we derived the black hole masses from the respective line widths ($\mathrm{H\alpha, H\beta, MgII}$) incorporating the line-dependent virial factors (Fig.\ref{fig3},\ref{fig4}b). Using the inferred line-widths from the $\cite{shen1}$ QSO catalog, we show the distribution of black hole masses as estimated from their respective broad emission lines, and compare them with the fiducial mass estimates from the catalog. In Fig.\ref{fig3}, the lower panel shows the spread in the T$\mathrm{_{BBB}}$ distribution and a similar comparison with the T$\mathrm{_{BBB}}$ derived from the fiducial mass estimates. For all the cases, we have used the mass accretion rate ($\mathrm{\dot{M}}$) from the QSO catalog and the accretion efficiency, $\mathrm{\eta} = 0.08\overline{3}$.

\section*{A semi-automatic spectral fitting procedure}
\begin{figure}%
\centering
    \subfloat{{\includegraphics[width=6cm, height = 5cm]{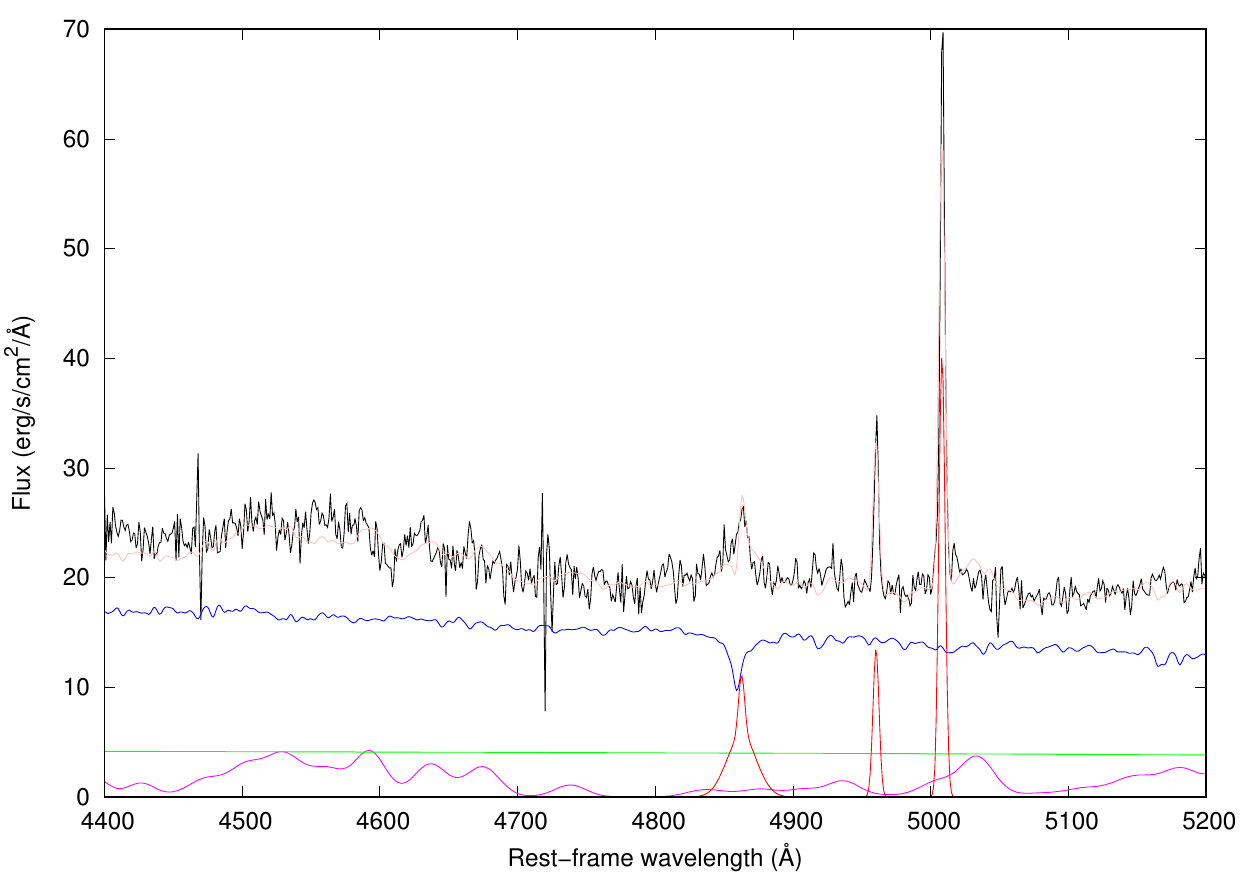} }}%
    \qquad
    \subfloat{{\includegraphics[width=6cm, height = 5cm]{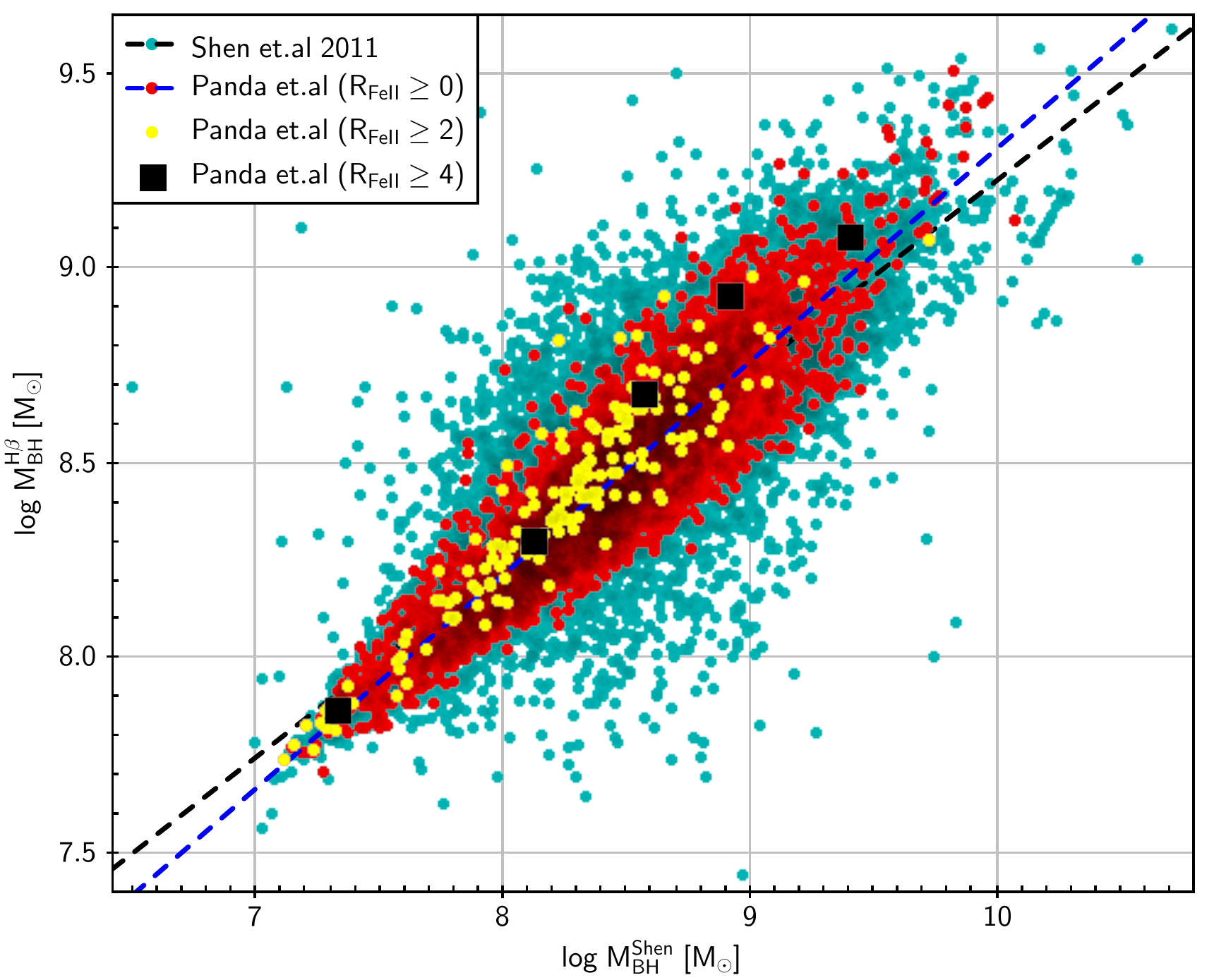} }}%
    \caption{(a) The expanded region of the H$\beta$ line of the source (pink line). Fit components: total flux (black line), starlight (blue line), power law (green dashed line), total line contribution (red line), and lower magenta line shows shape of the Fe II template d11-5-m20-20-5-mod broadened with a Gaussian of 700 km s$^{-1}$; (b) Comparing $\mathrm{M_{BH}}$ distribution: $\mathrm{M_{BH}}$ estimates from the $\cite{mej}$ prescription against the fiducial $\mathrm{M_{BH}}$ from $\cite{shen1}$ QSO catalog. The linear fits in black (QSO catalog: m(slope)=0.49, r=0.85) and blue (our sample: m=0.54, r=0.91)}%
    \label{fig4}%
\end{figure}

\vspace{-0.25cm}
Using the approach from $\cite{snow}$, we show a detailed modeling of the quasar spectrum which includes the Fe II pseudo-continuum (Fig.\ref{fig4}a). Owing to our fitting procedure that takes into account the Balmer continuum, starlight contamination and several FeII templates, the value of R$\mathrm{_{FeII}}$ thus estimated, is 4.045. This value quoted in the $\cite{shen1}$ QSO catalog is 6.56. Sources with strong Fe II emission are very difficult to model reliably, particularly the determination of the stellar content and the [OIII] line intensity is difficult due to strong Fe II contamination. These issues require careful analysis making a fully-automatic procedure relatively unreliable.

\section*{Discussions}
\vspace{-0.25cm}
We are currently working to automate this procedure to incorporate the CIGALE SED modeling, the CLOUDY interpolation and the semi-automatic spectral fitting. This will allow to put finer constraints on the estimates of T$\mathrm{_{BBB}}$ and simultaneously provide a more reliable value to their R$\mathrm{_{FeII}}$. One of our immediate goals is to see the distribution of these derived values of R$\mathrm{_{FeII}}$ against those in the catalog. With availability of newer and more reliable data (especially in the radio -- to study the distribution of RL/RQ in the context of these extreme objects, and X-rays - to model the optical-UV bump more accurately) the modeled SED structure will be refined. We will also test the line-dependent mass distribution replacing FWHM with a relatively better line-width parameter - the line dispersion, $\sigma_{line}$ $\cite{raf}$. We would like to test our CLOUDY generated SED that considers emission from a constant density single cloud (Panda et al. in prep) to these SEDs.

\section*{Acknowledgements}
\vspace{-0.25cm}
This conference has been organized with the support of the
Department of Physics and Astronomy ``Galileo Galilei'', the 
University of Padova, the National Institute of Astrophysics 
INAF, the Padova Planetarium, and the RadioNet consortium. 
RadioNet has received funding from the European Union's
Horizon 2020 research and innovation programme under 
grant agreement No~730562. The authors would like to acknowledge the financial support by Polish grant Nr. 2015/17/B/ST9/03436/.

\end{document}